\documentclass[a4paper]{jpconf}

\usepackage{graphicx}
\usepackage{amssymb}
\usepackage{hhline}
\usepackage{epsfig}
\usepackage{amsmath}
\usepackage{dcolumn}
\usepackage{multirow}
\usepackage{array}
\usepackage{color}

\newcommand{\be}{\begin{equation}}
\newcommand{\ee}[1]{\label{#1} \end{equation}}
\newcommand{\ba}{\begin{eqnarray}}
\newcommand{\ea}[1]{\label{#1} \end{eqnarray}}
\newcommand{\nl}{\nonumber \\}

\begin{document}
\title{A `soft + hard' model for heavy-ion collisions}

\author{K Urmossy$^{\,a}$, G G Barnaf\"oldi$^{\,a}$, Sz Harangoz\'o$^{\,a,b}$, T S Bir\'o$^{\,a}$,\\ and Z Xu$^{\,c}$}

\address{$^a$ Wigner Research Center for Physics of the Hungarian Academy of Sciences\nl29--33 Konkoly--Thege Mikl\'os Str., H-1121 Budapest, Hungary\vspace{1mm}}

\address{$^b$ E\"otv\"os University, 1/a P\'azm\'any P\'eter Walkway, H-1117 Budapest, Hungary\vspace{1mm}}

\address{$^c$ Physics Department, Brookhaven National Laboratory, Upton, NY 11973, USA}
\ead{karoly.uermoessy@cern.ch}

\begin{abstract}
We describe the spectra of neutral pions stemming from AuAu collisions at $\sqrt s$ = 200 AGeV in a `soft + hard' model. The model is based on the assumptions that hadrons stemming from hard processes are described via perturbative quantum chromodynamics improved parton model calculations, while those stemming from the Quark-Gluon Plasma (we refer to as soft yields) can be described in a super-statistical model induced by multiplicity fluctuations. The obtained dependence of the parameters of the model on the event centrality is compared to what is observed in PbPb collisions at $\sqrt s$ = 2.76 ATeV. 
\end{abstract}

\section{Introduction}
\label{sec:intro}

Among others, the phenomenon of jet-quenching supports that in ultra-relativistic heavy-ion collisions (AA) a new hadron source the Quark-Gluon Plasma (QGP) emerges besides the hard processes present in proton-proton (pp) and proton-nucleus (pA) collisions. We refer to hadrons stemming from these two types of sources as `soft' and `hard' yields. While perturbative quantum-chromodynamics improved parton model (pQCD) calculations  \cite{bib:pQCD,bib:pQCDGusty,bib:BGG_FF} describe the spectra of hadrons stemming from hard processes (pp, pA collisions and the high transverse momentum $p_T\gtrsim$ 6 GeV/c part of the spectra in AA collisions), pQCD methods fail to describe the $p_T\lesssim$ 6 GeV/c part of hadron spectra in AA collisions.

Disentangling soft and hard yields is a cumbersome problem for there may be low-$p_T$ hadrons stemming from hard processes as well as moderately high-$p_T$ ones created by soft processes \cite{bib:UKshLHC1}. Another difficulty is that due to the break-down of perturbation theory at low $p_T$, we might not trust pQCD results at low momenta. In this paper, we demonstrate two possible ways of obtaining the low-$p_T$ part of the hard hadron yields: $(i),$ we simply take the pQCD results; $(ii),$ we use a function fitted to the high-$p_T$ part of the pQCD spectra to make an extrapolation to low $p_T$. In the second case, we use a cut-power law function for the extrapolation, as such function turned out to provide an adequate description for hard hadron yields produced in pp collisions \cite{bib:Wong1}--\cite{bib:Wibig3}.

Once hard hadron yields (obtained in the $(i)$ and $(ii)$ scenarios) are subtracted from the measured transverse hadron spectra, we describe the remaining soft yields in the often used super-statistical framework \cite{bib:UKppNdep,bib:Wilk3,bib:Wilk4,bib:Wilk5,bib:Wilk6,bib:Wilk7,bib:BiroNfluktAA,bib:UKppFF,bib:UKeeFF}. In this paper, we use the interpretation in which, the cut-power law shape of the soft hadron distribution originates from multiplicity fluctuations rather than from non-extensive entropy or transport \cite{bib:BiroEPJA40,bib:BiroJako}.

The next section contains fits to neutral pion spectra measured in various centrality AuAu collisions at $\sqrt s$ = 200 AGeV by the PHENIX Collaboration \cite{bib:PHE}. To draw conclusions, we also discuss results obtained examining ALICE and CMS measurements \cite{bib:UKshLHC1}.

We note that similar soft and hard type separation of hadron yields has been suggested in models \cite{bib:Tang1,bib:Tang2,bib:Shao} in which, spectra measured in $pp$ collisions have been used as hard yields, and it has been conjectured that hard yields are suppressed at low $p_T$. Similar analysis' of the soft part of hadron spectra without the subtraction of hard yields are also presented in the literature \cite{bib:Wilk2}--\cite{bib:Wilk7}. 

\begin{figure}
\begin{center}
\includegraphics[width=0.45\textwidth]{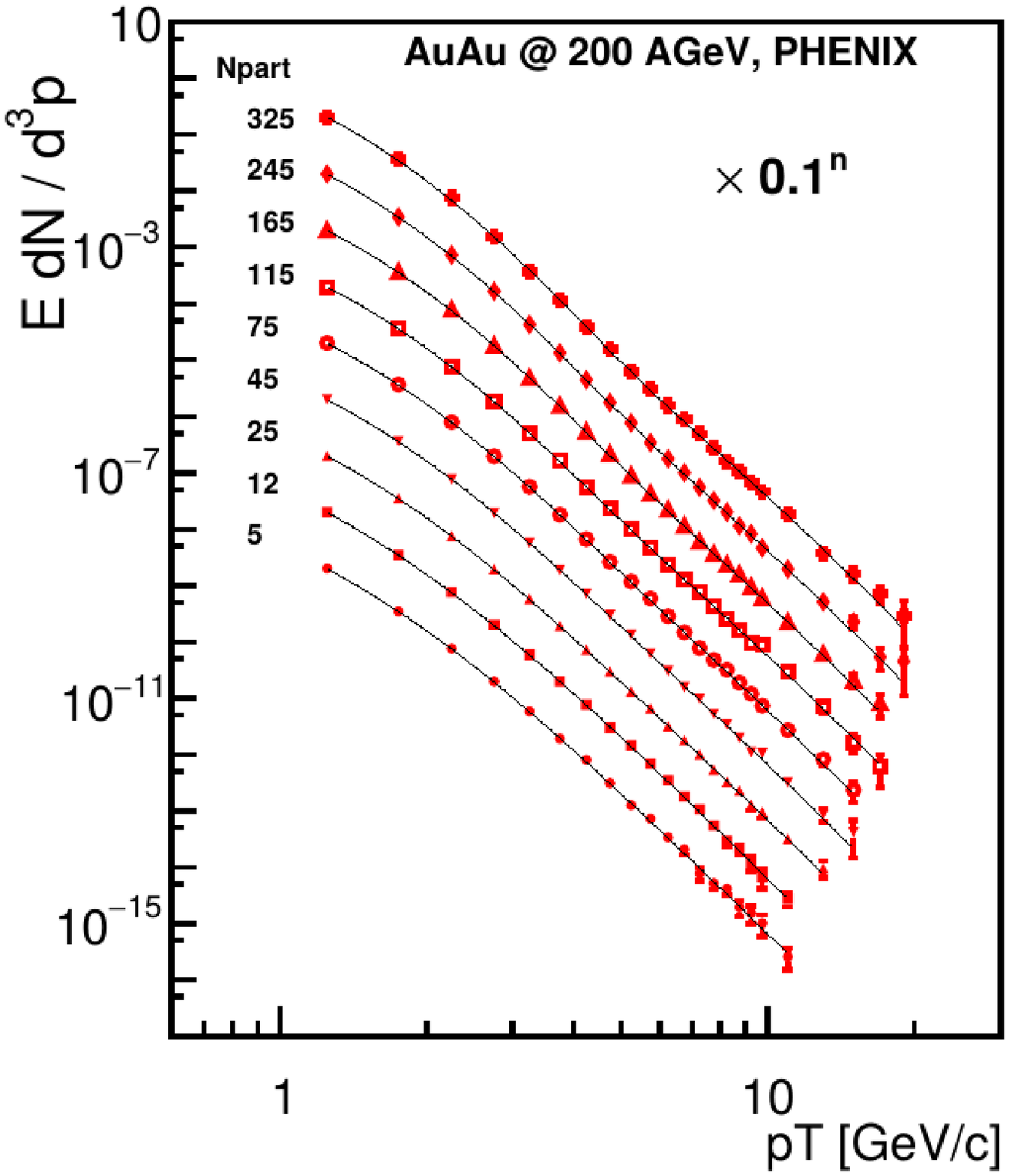}
\includegraphics[width=0.45\textwidth,height=0.53\textwidth]{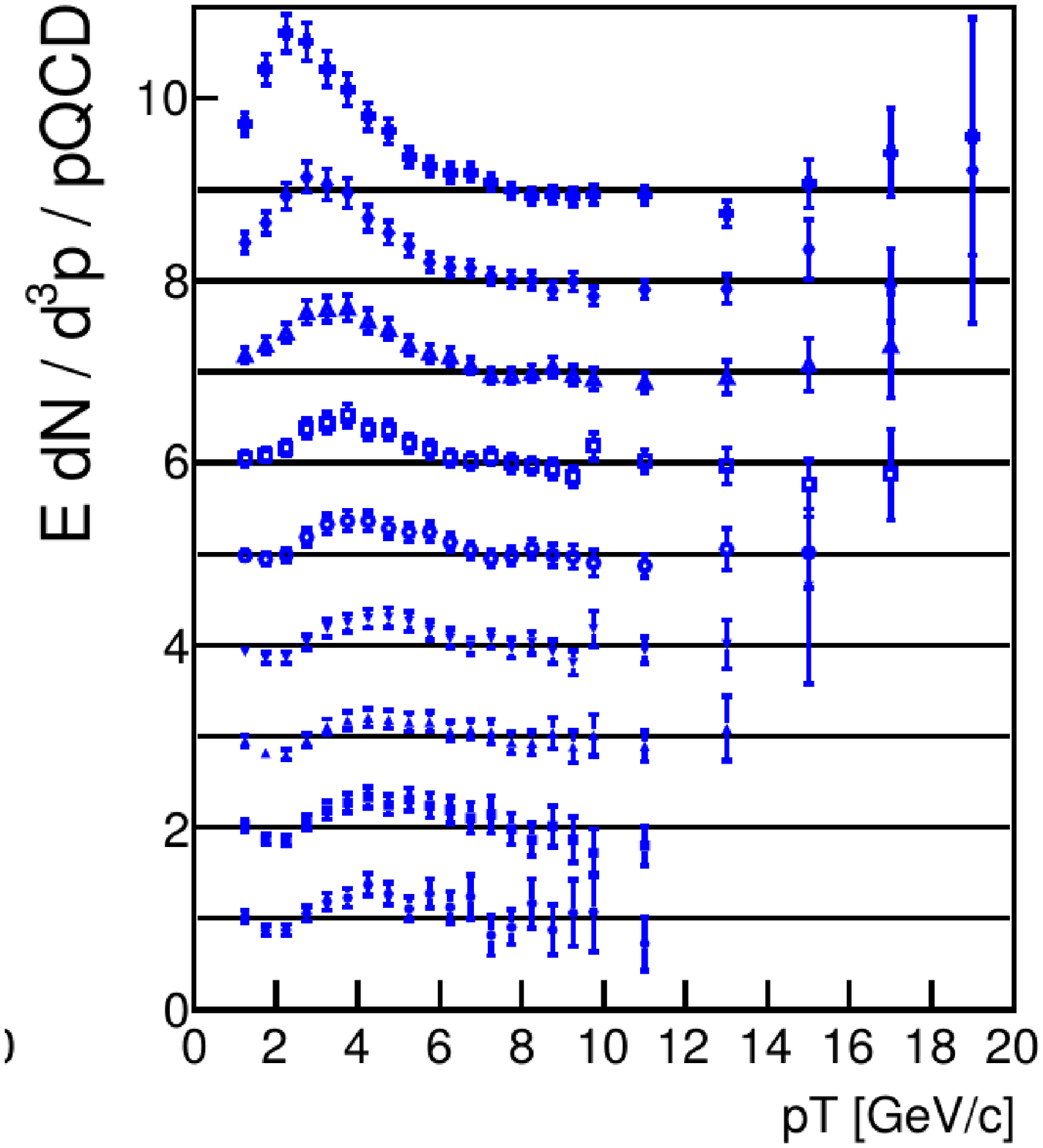}
\includegraphics[width=0.45\textwidth]{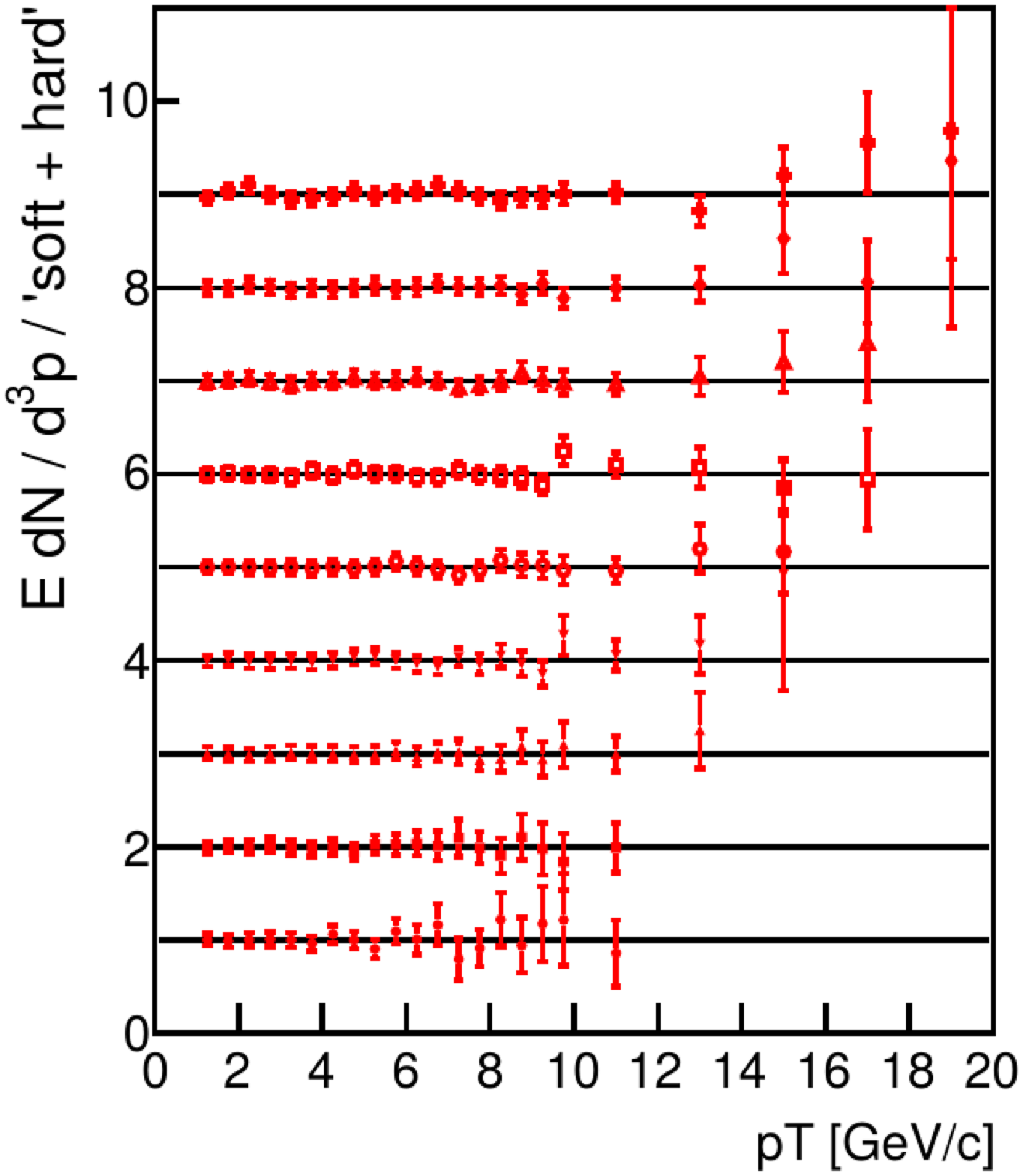}
\includegraphics[width=0.45\textwidth]{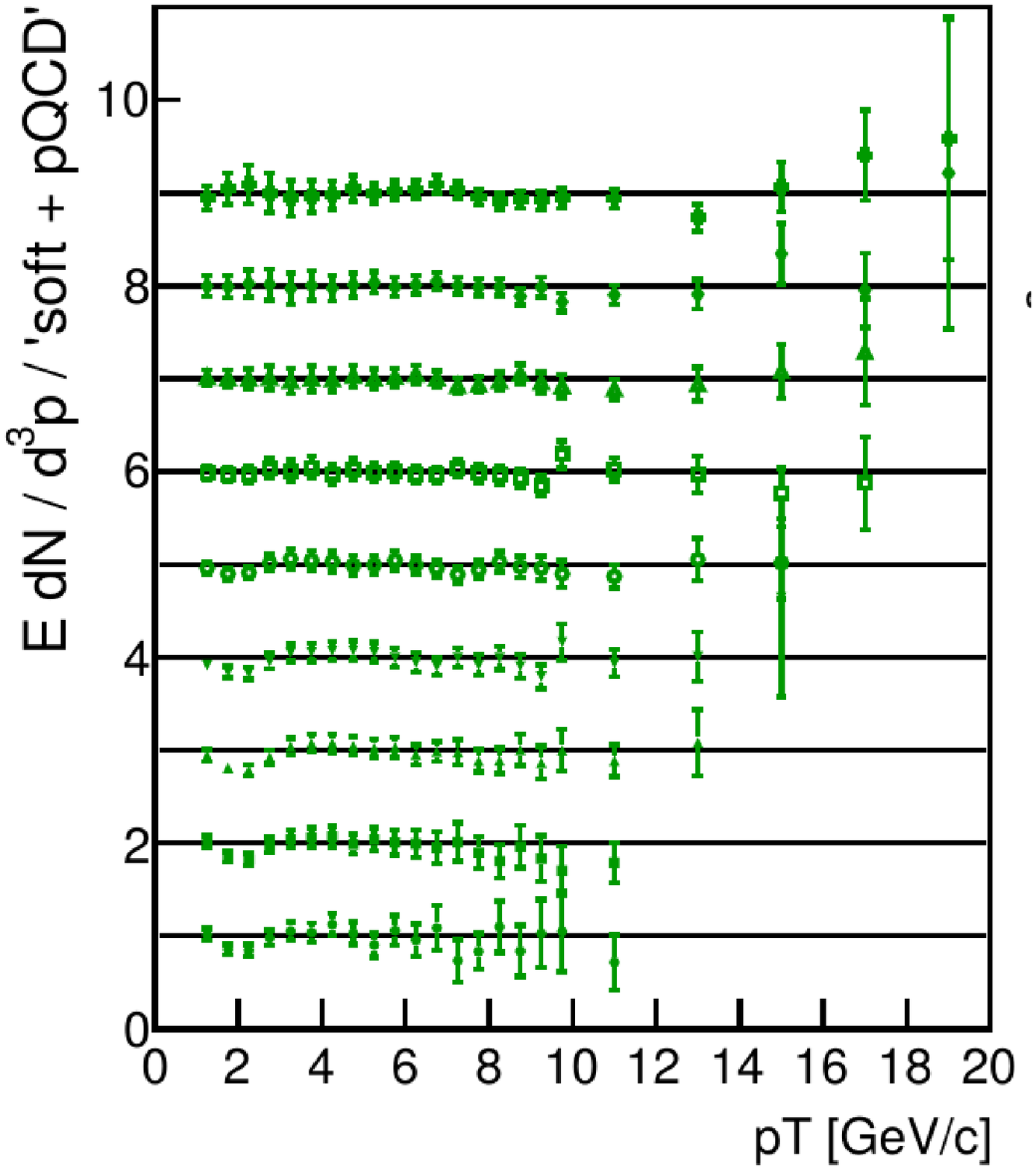}
\end{center}
\caption{{\bf Top-left,} transverse spectra of neutral pions stemming from various centrality AuAu collisions at $\sqrt s$ = 200 AGeV \cite{bib:PHE}. Topmost graphs are central events, bottommost graphs are peripheral events in all panels. Solid curves: fits of Eqs.~(\ref{eq12})--(\ref{eq13}), in which, both the soft and hard yields are fitted cut-power law functions as described in the $(i)$ scenario. {\bf Bottom-left,} Ratio of measured data and fits in the $(i)$ scenario. {\bf Top-right,} ratio of measured yields and pQCD results. {\bf Bottom-right,} ratio of measured yields and model results, in which, the hard yields are pQCD results and the soft yields are fitted cut-power law functions as described in the $(ii)$ scenario.\label{fig:dNdpT}}
\end{figure}

\begin{figure}
\begin{center}
\includegraphics[height=0.395\textheight]{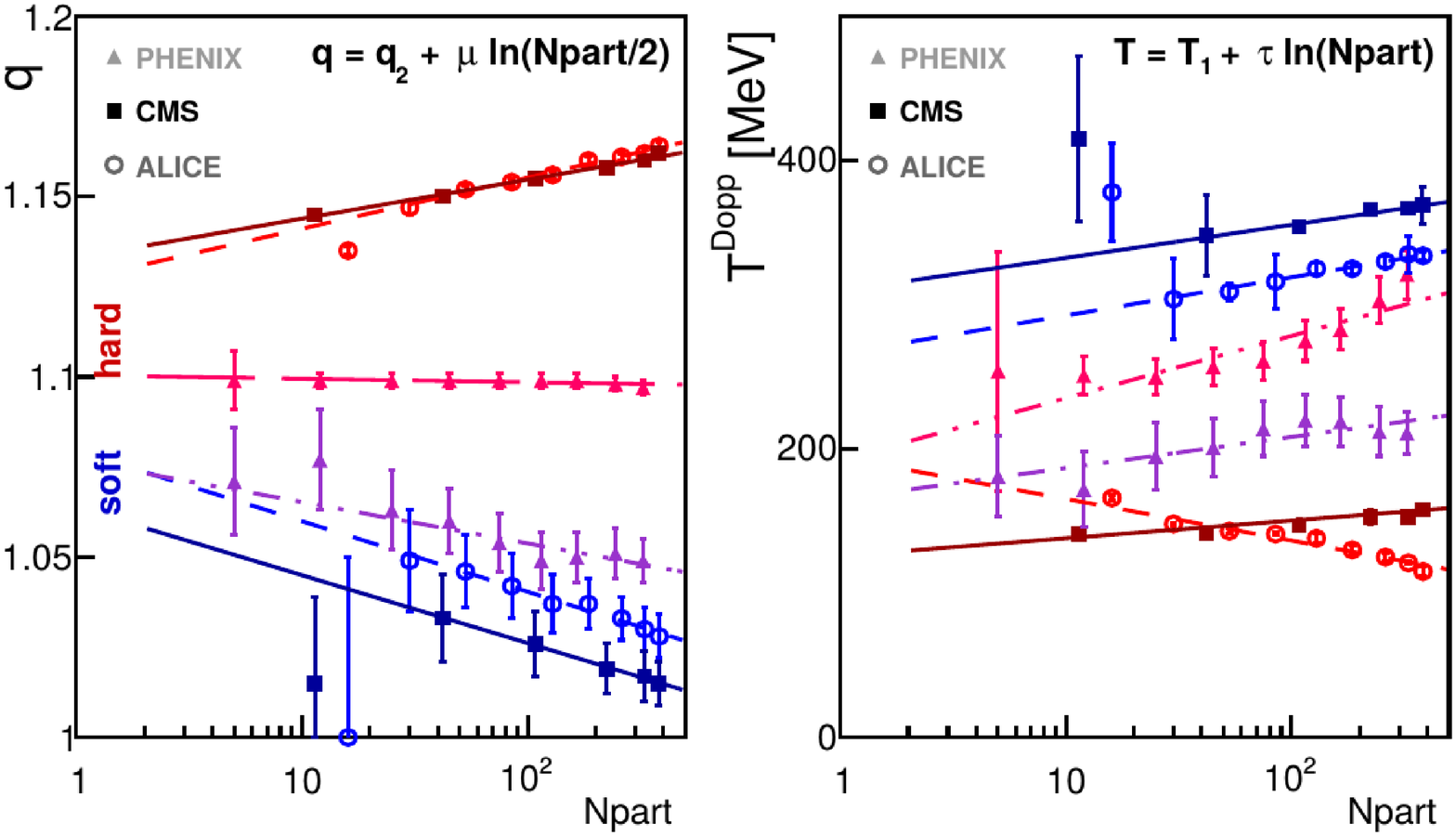}
\end{center}
\caption{Fitted parameters of Eqs.~(\ref{eq12})~--~(\ref{eq13}) in the $(i)$ scenario to PHENIX data on the spectra of $\pi^0$-s stemming from AuAu collisions at $\sqrt s$ = 200 AGeV (see Fig.~\ref{fig:dNdpT}). For comparison, results of fits to charged hadron spectra in PbPb collisions at $\sqrt s$ = 2.76 ATeV measured by the CMS and ALICE collaborations are also shown (see \cite{bib:UKshLHC1}). In both panels, data points belonging to soft hadron yields are colored in shades of blue, whereas those belonging to hard yields are reddish. Curves are fits of Eq.~(\ref{eq15}).\label{fig:params}}
\end{figure}

\section{Transverse spectrum in the `soft + hard' model}
\label{sec:spec}

As argued in Sec.~\ref{sec:intro} and in \cite{bib:UKshLHC1}, we make out the transverse spectrum of hadrons from the sum of hard and soft yields
\be
\left. \frac{dN}{2\pi p_T dp_T dy} \right|_{y=0} = f_{hard} + f_{soft}\;.
\ee{eq12}
As can be seen in the top-right panel of Fig.~\ref{fig:dNdpT}, pQCD results describe transverse spectra of neutral pions stemming from various centrality AuAu collisions at $\sqrt s$ = 200 AGeV for $p_T\geq$ 6 GeV/c. In the following, we examine two scenarios:
\begin{enumerate}
 \item[$i,$] We use pQCD results as hard yields for $p_T\leq$ 6 GeV/c too, and conjecture that the soft yields can be described by a cut-power law function.
  \item[$ii,$] We fit pQCD results for $p_T\geq$ 6 GeV/c with one cut-power law. We use this fitted function to extrapolate the hard yields to low $p_T$. Finally, we use a cut-power law with different parameters to fit the soft part of the spectra.
\end{enumerate}
In this approach, both the $p_T\leq$ 6 GeV/c and the $p_T\geq$ 6 GeV/c regions contain soft as well as hard yields. The cut-power law finctions are 
\be
f_i = A_i \left[1 + \frac{(q_i-1)}{T_i}[\gamma_i(m_T-v_i p_T) - m] \right]^{-1/(q_i-1)}
\ee{eq13}
where $i$ = soft or hard, $m = m_{\pi^0}$. The transverse mass is $m_T = \sqrt{p_T^2 + m^2}$, and $v_i$ is the isotropic part of the transverse flow with $\gamma_i = 1/\sqrt{1 - v^2_i}$ (see \cite{bib:UKshLHC1} for a detailed discussion). These yields have maxima at $p^{max}_{T} = \gamma_i\, m\, v_i$. As long as these maxima are below the measurement range, which is the case in this analysis, the transverse flow, $v_i$ cannot be determined accurately. As in this paper we analyse neutral pions, the arguments of $f_i$-s may be approximated by $[\gamma_i(m_T - v_i p_T) -m]/T_i \approx p_T/T^{Dopp}_i$ with the Doppler-shifted parameters
\be
T^{Dopp}_i = T_i\,\sqrt{\frac{1+v_i}{1-v_i}} \;.
\ee{eq14}

\section*{Results in the $(i)$ scenario}

Fit results of Eqs.~(\ref{eq12})--(\ref{eq13}) in the two scenarios explained above, are shown in Fig.~\ref{fig:dNdpT}. The dependence of the obtained $q$ and $T^{Dopp}$ parameters of the soft and hard yields on the event centrality (number of participating nucleons $N_{part}$) can be fitted by (see Fig.~\ref{fig:params})
\ba
q_i &=& q_{2,\,i} + \mu_i \ln(N_{part}/2) \;,\nl
T^{Dopp}_i &=& T_{1,\,i} + \tau_i \ln(N_{part}) \;.
\ea{eq15}
Though the actual value of the transverse flow velocity cannot be determined in this model from the spectra of charged hadrons, it may be guessed using the value of the QGP-hadronic matter transition temperature obtained from lattice-QCD calculations. As the values of fitted $T^{Dopp}_{soft}$ scatter around $\sim$200 MeV at RHIC (and $\sim$340 MeV at LHC), in case of a flow velocity of $v_{soft} \approx$  0.5 at RHIC (and $v_{soft} \approx$  0.6 at LHC), the real $T_{soft}$ values would scatter around 170 MeV, which is close to the lattice result obtained e.g. in \cite{bib:lQCD}.

$q_{soft}$ and $T_{soft}$ may have statistical meaning. If in each nucleus-nucleus collision, hadrons stemming from soft processes have either Boltzmann--Gibbs or micro-canonical distribution, however, hadron multiplicity fluctuates according to the negative-binomial distribution, then the cut-power law function Eq.~(\ref{eq13}) emerges as the average hadron spectrum \cite{bib:UKppNdep,bib:Wilk3,bib:Wilk4,bib:Wilk5,bib:Wilk6,bib:Wilk7,bib:BiroNfluktAA,bib:UKppFF,bib:UKeeFF}. If $q\rightarrow 1$, Eq.~(\ref{eq13}) tends to the Boltzmann distribution, while the multiplicity distribution tends to the Poissonian distribution. Such tendency can be seen in the left panel of Fig.~\ref{fig:params}. We may interpret this phenomenon as the soft yields are closer to the thermal, if either $N_{part}$ or $\sqrt s$ grows. As can be seen in the top panels of Fig.~\ref{fig:dNdpT} and right panel of Fig.~\ref{fig:params}, the more central the collision, the bigger and hotter the soft yield.

The behaviour of the parameters of the hard yields have to be determined by scale evolution equations of QCD in yet unexamined ways. Nevertheless, it can be seen that as $\sqrt s$ increases, $q_{hard}$ increases as well, while the dimensionless parameter $T_{hard}/\sqrt{s}$ appearing in the cut-power-law-type fragmentation functions introduced in \cite{bib:BGG_FF}, decreases. These trends seen in Fig.~\ref{fig:params} are similar to what have been found in \cite{bib:BGG_FF}. 

\begin{table}[h!]
\begin{center}
\begin{tabular}{ c|*{4}l }
                      & $q_{2,soft}$         & $q_{2,hard}$         & $\mu_{soft}$          & $\mu_{hard}$            \\ \hhline{=|====}  \noalign{\smallskip}
CMS                   & 1.058 $\pm$ 0.025    & 1.136 $\pm$ 0.001    & -0.008 $\pm$ 0.005    & 0.005 $\pm$ 0.0003      \\ \hline  \noalign{\smallskip}
ALICE                 & 1.074 $\pm$ 0.018    & 1.131 $\pm$ 0.002    & -0.009 $\pm$ 0.004    & 0.006 $\pm$ 0.0006      \\ \hline  \noalign{\smallskip}
PHENIX                & 1.073 $\pm$ 0.016    & 1.100 $\pm$ 0.002    & -0.005 $\pm$ 0.004    & 0.000 $\pm$ 0.0006      \\ \hline  \noalign{\bigskip}
                      & $T^{soft}_{1}$ [MeV] & $T^{hard}_{1}$ [MeV] & $\tau_{soft}$ [MeV]   & $\tau_{hard}$ [MeV]     \\ \hhline{=|====}  \noalign{\smallskip}
CMS                   & 310 $\pm$ 20         & 126 $\pm$ 5          & $\;\;$9.9 $\pm$ 3.7   & $\;\;\;$5.3 $\pm$ 0.8   \\ \hline  \noalign{\smallskip}
ALICE                 & 266 $\pm$ 16         & 194 $\pm$ 2          & 11.5 $\pm$ 2.9        & -12.5 $\pm$ 0.5         \\ \hline  \noalign{\smallskip}
PHENIX                & 165 $\pm$ 26         & 192 $\pm$ 20         & $\;\;$9.3  $\pm$ 5.5  & $\;$18.7  $\pm$ 4.6   \\ \hline  \noalign{\smallskip}
\end{tabular} 
\caption{Parameters of Eq.~(\ref{eq15}) obtained from fits shown in Fig.~\ref{fig:params}.\label{tab:qTfit}}
\end{center}
\end{table}

\section*{Results in the $(ii)$ scenario}

In the second scenario, the hard yields were taken directly from the pQCD calculations, and the soft yields were fitted by the cut-power law function Eq.~(\ref{eq13}) (see right panels of Fig.~\ref{fig:dNdpT}). The obtained soft distributions were practically exponentials, thus we may conclude that if pQCD results \cite{bib:pQCD,bib:pQCDGusty} are trustworthy at $p_T\leq$ 6 GeV/c, the soft yields are likely to be thermal. This scenario has been thoroughly discussed in the literature \cite{bib:OldBoys1}--\cite{bib:Liu1}.

\section*{Acknowledgement}
\label{sec:ack}
This work was supported by Hungarian OTKA grants K104260, NK106119, and
NIH TET 12 CN-1-2012-0016. Author GGB also thanks the J\'anos Bolyai
Research Scholarship of the Hungarian Academy of Sciences.

\end{document}